\documentclass[final,3p,times,twocolumn]{elsarticle}
                                   
\usepackage{graphicx}
\usepackage{amssymb}
\usepackage{amsmath}
\usepackage{hyperref} 
\usepackage{longtable}
\biboptions{sort&compress}
\usepackage{xcolor}

\journal{Physics Letters B}

\newcommand{\preprint}{
 \setlength{\unitlength}{1mm}{\hbox{\begin{picture}(0,0)
      \put(160,10){\mbox{\footnotesize%
        ADP-10-25/T721}}\end{picture}}}}

\begin{document}

\begin{frontmatter}

\title{\preprint  Roper Resonance in 2+1 Flavor QCD}

\author[adl]{M. Selim Mahbub}
\author[adl]{Waseem Kamleh}
\author[adl]{Derek B. Leinweber}
\author[adl,csiro]{Peter J. Moran}
\author[adl]{Anthony G. Williams}
\address[adl]{Special Research Centre for the Subatomic Structure of Matter, Adelaide, South Australia 5005, Australia, \\ and Department of Physics, University of Adelaide, South Australia 5005, Australia.}
\address[csiro]{CSIRO Mathematics, Informatics and Statistics, Private Bag 33, Clayton South, VIC 3169, Australia.}

\begin{abstract}

The low-lying even-parity states of the nucleon are explored in
lattice QCD using the PACS-CS collaboration 2+1-flavor dynamical-QCD
gauge-field configurations made available through the International
Lattice Datagrid (ILDG).
The established correlation-matrix approach is used, in which
various fermion source and sink smearings are utilized to provide an
effective basis of interpolating fields to span the space of
low-lying energy eigenstates.
Of particular interest is the nature of the first excited state of
the nucleon, the $N{\frac{1}{2}}^{+}$ Roper resonance of $P_{11}$
pion-nucleon scattering.
The Roper state of the present analysis approaches the
physical mass, displaying significant chiral curvature at the
lightest quark mass.
These full QCD results, providing the world's first insight into the
nucleon mass spectrum in the light-quark regime, are significantly
different from those of quenched QCD and provide interesting insights
into the dynamics of QCD.
\end{abstract}

\begin{keyword}
 Roper resonance \sep dynamical fermions \sep Lattice QCD

 \PACS 11.15.Ha \sep 12.38.Gc \sep 12.38.-t \sep 13.75.Gx

\end{keyword}

\end{frontmatter}

\section{Introduction}
The first positive parity resonance of the nucleon, the
$N{\frac{1}{2}}^+(1440)$ or Roper resonance, has been the
subject of extensive interest since its discovery in 1964 
\cite{Roper:1964zz}.
This $P$-wave isospin-1/2 spin-1/2 ($P_{11}$) pion-nucleon resonance
has held the curiosity and imagination of the nuclear and particle
physics community due to its surprisingly low mass.
For example, in constituent quark models the lowest-lying odd-parity
state occurs below the ${P}_{11}$ state
\cite{Isgur:1977ef,Isgur:1978wd}
whereas in Nature, the negative parity $N{\frac{1}{2}}^{-}(1535)$
${\rm S}_{11}$ state is almost 100 MeV {\it above} the Roper
resonance.

This phenomenon has led to wide speculation on the possible exotic
nature of the Roper resonance.  For example, the Roper resonance has
been described as a hybrid baryon state with explicitly excited gluon
field configurations~\cite{Li:1991yba,Carlson:1991tg}, as a
breathing mode of the ground state~\cite{Guichon:1985ny} or a state
which can be described in terms of a five quark (meson-baryon)
state~\cite{Krehl:1999km}.

The elusive nature of this low-lying resonance is not constrained to
model calculations alone.  There have been several investigations of
the low-lying nucleon spectrum using the first-principles approach of
lattice field theory.  

The lattice approach to Quantum Chromodynamics (QCD) provides a
non-perturbative tool to explore the properties of hadrons from the
first principles of this fundamental quantum field theory.
Numerical simulations of QCD on a space-time lattice with the light
up, down and strange dynamical-quark masses similar to those of Nature
are now possible \cite{Aoki:2008sm}.  As such, some long-standing
problems in nuclear-particle physics are now being resolved.  For
example, the ground-state hadron spectrum is now well
understood~\cite{Durr:2008zz}.

However, gaining knowledge of the excited-state spectrum presents
additional challenges.  The Euclidean-time correlation function
provides access to a tower of energy eigenstates in the form of a sum
of decaying exponentials with the masses of the states in the
exponents. The ground state mass, being the lowest energy state, has
the slowest decay rate, and is obtained through the analysis of the
large-time behaviour of this function.  However, the excited states appear in the
sub-leading exponentials. Extracting excited state masses from these
exponents is intricate as the correlation functions decay
quickly and the signal to noise ratio deteriorates rapidly.
In addition, the spectrum is composed of both single-particle states
and multiple-particle states interacting and mixing in the finite
physical volume of the lattice.  Understanding the finite-volume
dependence of these states and linking them to the resonances of
Nature is a long-term program of the lattice QCD community.

In this letter we report the excited-state energy spectrum of the nucleon
in the light quark-mass regime of QCD for the first time.  Of
particular note is the identification of a new low-lying state
associated with the Roper resonance of Nature.

Several attempts have been made in the past to find the elusive
low-lying Roper state in the lattice
framework~\cite{Leinweber:1994nm,Lee:1998cx,Gockeler:2001db,Sasaki:2001nf,Melnitchouk:2002eg,Lee:2002gn,Edwards:2003cd,Mathur:2003zf,Sasaki:2003xc,Basak:2007kj,Bulava:2009jb,Bulava:2010yg,Engel:2010my}.
The results reported therein are as much about the development of
lattice techniques as they are about the nucleon spectrum.  Where the
lattice techniques are regarded as robust, a low-lying Roper state was
not observed.  The difficulties lie in finding effective methods
to isolate the energy eigenstates of QCD and in accessing the light
quark mass regime of QCD.

The `Variational method'~\cite{Michael:1985ne,Luscher:1990ck} is the
state-of-the-art approach for determining the excited state hadron
spectrum.  It is based on the creation of a matrix of correlation
functions in which different superpositions of excited state
contributions are linearly combined to isolate the energy eigenstates.
A diversity of excited-state superpositions is central to the success
of this method.

Early implementations of this method, using a variety of
standard spin-flavor interpolating fields of fixed source-distribution
size, were not successful in isolating energy eigenstates.  Instead the
putative eigenstates were superpositions of
energy-states~\cite{Mahbub:2009nr} and the low-lying Roper state was
hidden by excited-state contaminations.  A solution to this problem
was established in Refs.~\cite{Mahbub:2009aa,Mahbub:2010jz} where a
low-lying Roper state was isolated.  Key to this approach, used
herein, is the utilization of a diverse range of fermion source and
sink smearings in creating the matrix of correlation
functions.  The diversity of smearings leads to a wide variety of
superpositions of excited-state contributions, providing a suitable
basis for constructing linear combinations which isolate the
eigenstates.

These effective techniques
\cite{Mahbub:2009nr,Mahbub:2009aa,Mahbub:2010jz,Mahbub:2010me} were
developed in the quenched approximation and we bring them to the
dynamics of full QCD for the first time here.
The low-lying even-parity states of the nucleon are explored in
full QCD using 2+1-flavor dynamical-QCD
gauge-field configurations \cite{Aoki:2008sm}.
Whereas other recent full QCD
analyses~\cite{Bulava:2009jb,Bulava:2010yg,Engel:2010my} report a
first positive parity excited state that appears too high to be
considered as the Roper resonance, we will illustrate how the
low-lying Roper state of the present analysis approaches the physical
mass of Nature, displaying significant chiral curvature at the
lightest quark mass (corresponding to a pion mass of 156 MeV, only 
slightly above the physical value of 140 MeV).
These full QCD results, providing the world's first insight into the
nucleon mass spectrum in the light-quark regime, are significantly
different from those of quenched QCD and provide interesting insights
into the dynamics of QCD.

In constructing our correlation matrix for the nucleon spectrum, we
consider the two-point correlation function matrix with momentum
$\vec{p} = 0$
\begin{align}
G_{ij}^{\pm}(t) &= \sum_{\vec x}\, {\rm Tr}_{\rm sp}\, \{
\Gamma_{\pm}\, \langle\Omega\vert\chi_{i}(x)\, \bar\chi_{j}(0)\vert\Omega\rangle\}, \\
          &=\sum_{\alpha} \, \lambda_{i}^{\alpha}\,
\bar\lambda_{j}^{\alpha} \, e^{-m_{\alpha}t},
\end{align}
where Dirac indices are implicit.  Here, $\lambda_{i}^{\alpha}$ and
$\bar\lambda_{j}^{\alpha}$ are the couplings of the interpolators
$\chi_{i}$ and $\bar\chi_{j}$ at the sink and source respectively and
$\alpha$ enumerates the energy eigenstates with mass
$m_{\alpha}$. $\Gamma_{\pm}=\frac{1}{2}(\gamma_{0}\pm 1)$ projects the
parity of the eigenstates.

Since the only $t$ dependence comes from the exponential term, one can
seek a linear superposition of interpolators,
${\bar\chi}_{j}u_{j}^{\alpha}$, such that
\begin{align}
G_{ij}(t_{0}+\triangle t)\, u_{j}^{\alpha} & = e^{-m_{\alpha}\triangle
  t}\, G_{ij}(t_{0})\, u_{j}^{\alpha},
\end{align}  
for sufficiently large $t_{0}$ and $t_{0}+\triangle t$.  Multiplying
the above equation by $[G_{ij}(t_{0})]^{-1}$ from the left leads to an
eigenvalue equation
\begin{align}
[(G(t_{0}))^{-1}\, G(t_{0}+\triangle t)]_{ij}\, u^{\alpha}_{j} & = c^{\alpha}\, u^{\alpha}_{i},
 \label{eq:right_evalue_eq}
\end{align} 
where $c^{\alpha}=e^{-m_{\alpha}\triangle t}$ is the eigenvalue.
Similar to Eq.~(\ref{eq:right_evalue_eq}), one can also solve the left
eigenvalue equation to recover the $v^{\alpha}$ eigenvector
\begin{align}
v^{\alpha}_{i}\, [G(t_{0}+\triangle t)\, (G(t_{0}))^{-1}]_{ij} & =
c^{\alpha}\, v^{\alpha}_{j}.
\label{eq:left_evalue_eq}
\end{align} 
The vectors $u_{j}^{\alpha}$ and $v_{i}^{\alpha}$  diagonalize the
correlation matrix at time $t_{0}$ and $t_{0}+\triangle t$ and provide
the projected correlator
\begin{align}
v_{i}^{\alpha}\, G_{ij}^{\pm}(t)\, u_{j}^{\beta} & \propto \delta^{\alpha\beta}.
\label{projected_cf} 
\end{align} 
The parity projected, eigenstate projected correlator 
\begin{align}
G^{\alpha}_{\pm}& \equiv v_{i}^{\alpha}\, G^{\pm}_{ij}(t)\, u_{j}^{\alpha} ,
 \label{projected_cf_final}
\end{align}
is then analyzed using standard techniques to obtain the masses of
different states
\cite{Melnitchouk:2002eg,Mahbub:2009nr,Blossier:2009kd}.

\begin{figure*}[!t]
  \begin{center} 
 $\begin{array}{l@{\hspace{-1cm}}c} 
  \null\hspace{-0.3cm}\includegraphics[height=0.37\textwidth]{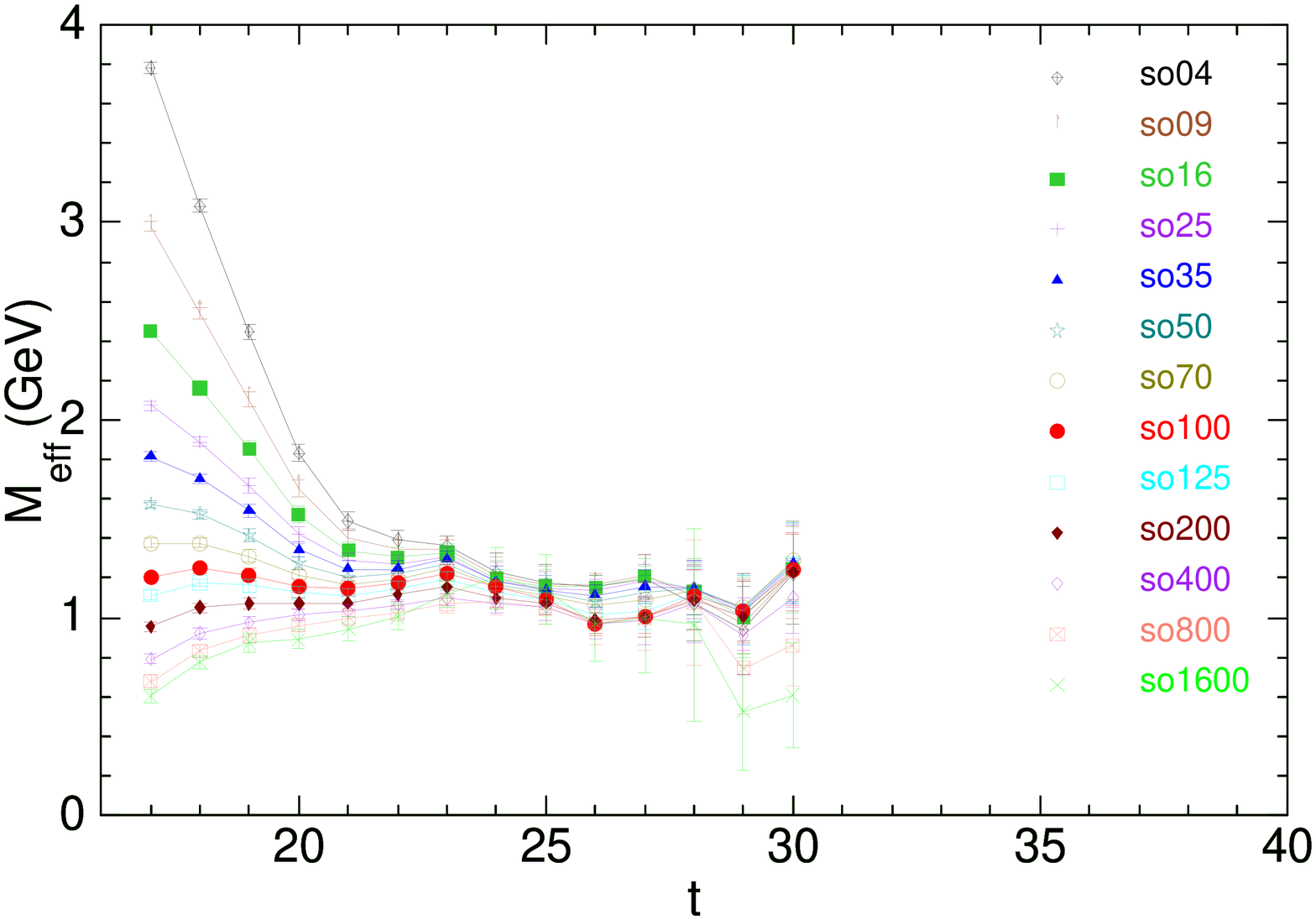} &
  \includegraphics[height=0.37\textwidth]{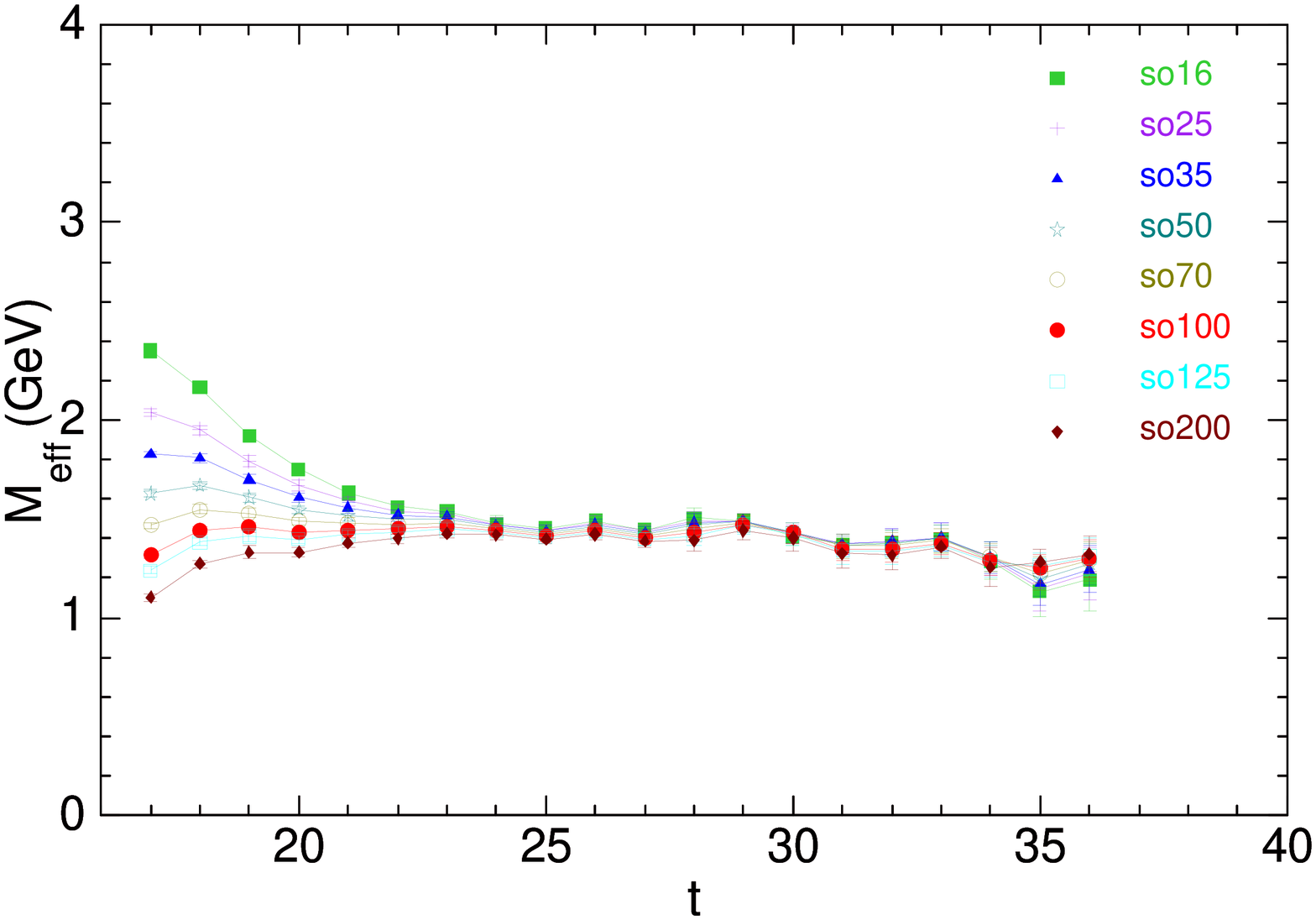} 
 \end{array}$
 \end{center}
\vspace{-0.6cm}
 \caption{(Color online).  Effective mass from smeared-source to
   point-sink correlators for various levels of smearings at the
   source for $\kappa_{ud}=0.13770$ (left) and $\kappa_{ud}=0.13700$
   (right), corresponding to pion masses of 293 and 702 MeV, respectively.}
 \label{fig:eff_Kud_013770_013700_n_so_allsmear_x1x1}
\vspace{-0.2cm}
\end{figure*}

The PACS-CS $2+1$ flavor dynamical-fermion
configurations~\cite{Aoki:2008sm} made available through the ILDG are
used herein.  These configurations use the
non-perturbatively ${\cal{O}}(a)$-improved Wilson fermion action and
the Iwasaki-gauge action~\cite{Iwasaki:1983ck}.  The lattice volume is
$32^{3}\times 64$, with $\beta=1.90$ providing a lattice spacing
$a=0.0907$ fm.
Five values of the (degenerate)
up and down quark masses are considered, with hopping parameter values of
$\kappa_{ud}=0.13700, 0.13727, 0.13754, 0.13770\text{ and }0.13781$,
corresponding to pion masses of $m_{\pi}$ = 0.702,
  0.572, 0.413, 0.293, 0.156 GeV~\cite{Aoki:2008sm}; for the strange quark
$\kappa_{s}=0.13640.$  
We consider 350 configurations for the four heavier quarks, and
198 configurations for the
lightest quark. An ensemble of 750 samples for the lightest quark mass
is created by using multiple fermion sources on each configuration,
spaced to sample approximately independent regimes of each configuration. 
Our error analysis is performed using a second-order jackknife method, where the
${\chi^{2}}/{\rm{dof}}$ for projected correlator fits is obtained via
a covariance matrix analysis.  Our fitting method is discussed in
Refs.~\cite{Mahbub:2010jz,Mahbub:2009nr}.

The complete set of local interpolating fields for the
spin-$\frac{1}{2}$ nucleon are considered herein. Three different
spin-flavor combinations of nucleon interpolators are considered, 
\begin{align}
\label{eq:interp_x1}
\chi_1(x) &= \epsilon^{abc}(u^{Ta}(x)C{\gamma_5}d^b(x))u^{c}(x)\, , \\
\label{eq:interp_x2}
\chi_2(x) &= \epsilon^{abc}(u^{Ta}(x)Cd^b(x)){\gamma_5}u^{c}(x)\, , \\
\label{eq:interp_x4}
\chi_4(x) &= \epsilon^{abc}(u^{Ta}(x)C{\gamma_5}{\gamma_4}d^b(x))u^{c}(x).
\end{align}

Each interpolator has a unique Dirac structure giving rise to
different spin-flavor combinations. Moreover, as each spinor has upper
and lower components, with the lower components containing an implicit
derivative, different combinations of zero and two-derivative
interpolators are provided. The local scalar-diquark nucleon
interpolator, $\chi_{1}$, is well known to have a good overlap with
the ground state of the nucleon. Also, this $\chi_{1}$ interpolator is
able to extract a low-lying Roper state in quenched
QCD~\cite{Mahbub:2010jz}. On the other hand, the $\chi_{2}$
interpolator, which vanishes in the non-relativistic limit, couples
strongly to higher energy states. The interpolator $\chi_{4}$ is the
time component of the local spin-$\frac{3}{2}$ isospin-$\frac{1}{2}$
interpolator which also couples to spin-$\frac{1}{2}$ states.  It
provides a different linear combination of zero- and two-derivative
terms.

In constructing our correlation matrices, we first
consider an extensive sample of different levels of gauge-invariant
Gaussian smearing \cite{Gusken:1989qx} at the fermion source and
sink, including 4, 9, 16, 25, 35, 50, 70, 100, 125, 200, 400,
800 and 1600 sweeps.
These levels of smearing correspond to rms radii in lattice
units ($a \simeq 0.09$ fm) of 1.20, 1.79, 2.37, 2.96, 3.50, 4.19,
4.95, 5.920, 6.63, 8.55, 12.67, 15.47 and 16.00.  

Fig.\ref{fig:eff_Kud_013770_013700_n_so_allsmear_x1x1} displays
effective mass plots, $m(t) = \ln \{ G^+_{ij}(t) / G^+_{ij}(t+1) \}$
for smeared-source to point-sink correlators of $\chi_{1}$.  The variation in the
superposition of excited state contributions is revealed in the
different approaches of the effective mass to the ground state plateau
where the results converge.  From these plots, it is clear that the
correlation matrix analysis for excited state contributions will be
most effective in the regime $t=17-21$ where there is diversity in the
curves.

\begin{table*}
   \begin{center}
   \vspace{-0.4cm}
   \caption{Smearing levels used in constructing $4\times 4$
     correlation matrix bases.}
   \vspace{0.2cm}
   \label{table:varieties_of_4x4_matrices}
   \begin{tabular}{c|cccccccccc} 
   \hline
   Sweeps $\rightarrow$  & 16 & 25 & 35 & 50 & 70 & 100 & 125 & 200 & 400 & 800  \\
   \hline
   Basis No. $\downarrow$ & \multicolumn{10}{c}{Bases}  \\
   \hline 

1  & 16 & -  & 35 & -   & 70 & 100 & -   & -   & -   & - \\ 
2  & 16 & -  & 35 & -   & 70 & -   & 125 & -   & -   & - \\
3  & 16 & -  & 35 & -   & -  & 100 & -   & 200 & -   & - \\
4  & 16 & -  & 35 & -   & -  & 100 & -   & -   & 400 & - \\
5  & 16 & -  & -  & 50  & -  & 100 & 125 &  -  & -   & - \\
6  & 16 & -  & -  & 50  & -  & 100 & -   & 200 & -   & - \\
7  & 16 & -  & -  & 50  & -  & -   & 125 &  -  & -   & 800 \\
8  & -  & 25 & -  & 50  & -  & 100 & -   & 200 & -   & - \\
9  & -  & 25 & -  & 50  & -  & 100 & -   & -   & 400 & - \\
10 & -  & -  & 35 & -   & 70 & -   & 125 & -   & 400 & - \\ 
\hline 
 \end{tabular}
 \end{center}
\vspace{-0.6cm} 
\end{table*}

To explore the low-lying eigenstates of the nucleon spectrum, we
construct several $4\times 4$ correlation matrices of $\chi_{1}$ as described in
Table~\ref{table:varieties_of_4x4_matrices}.  These matrices provide
robust results for the lowest three energy eigenstates observed, with the
highest energy level accommodating the fourth eigen-energy and any
residual strength from higher states not eliminated via
Euclidean time evolution.

\begin{figure}[!b]
  \begin{center} 
  \includegraphics[height=0.46\textwidth,angle=90]{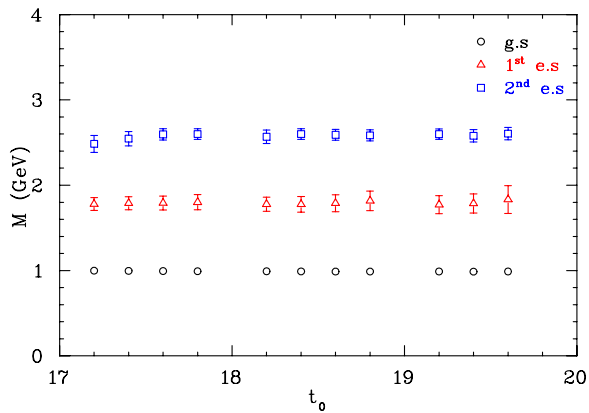}
 \end{center}
\vspace{-0.5cm}
 \caption{(Color online). Masses from the projected correlation
   functions as shown in Eq.~\ref{projected_cf_final}, for each set of
   variational parameters $t_{0}$ (major axis) and $\triangle t$
   (minor axis).  This figure corresponds to the lightest quark mass,
   $\kappa_{ud}=0.13781,$ for which $m_{\pi}=156\, \text{MeV}$, and the 3rd basis of
   Table~\ref{table:varieties_of_4x4_matrices}.}
 \label{fig:ma_Kud_013781_4x4_x1x1}
\end{figure}

The masses from the projected correlation functions obtained from the
correlation matrix analysis are very consistent over the variational
parameters ($t_{0}$,$\triangle t$) as illustrated in
Fig.~\ref{fig:ma_Kud_013781_4x4_x1x1}. 
Careful
examination of Fig.~\ref{fig:ma_Kud_013781_4x4_x1x1} reveals some
systematic drift in the second excited state mass at small $\triangle t$ for $t_{0}=17,18.$  This
emphasises the preference for selecting larger values of
($t_{0}$,$\triangle t$) as discussed in
Refs.~\cite{Mahbub:2009aa,Mahbub:2010jz,Blossier:2009kd}.  However, larger
uncertainties are evident for $t_{0}=18,19$ with large $\triangle t$ due to
the suppression of excited states via Euclidean time evolution.  
We select $t_{0}$=18, $\triangle t$=2 as providing the best balance
between these systematic and statistical uncertainties~\cite{Mahbub:2009nr,Mahbub:2010jz}.  These
variational parameters also provide projected correlation functions
having favorable $\chi^{2}/{\rm dof}$ in the effective mass
fits.

\begin{figure}[!b]
  \begin{center}
 \includegraphics[height=0.46\textwidth,angle=90]{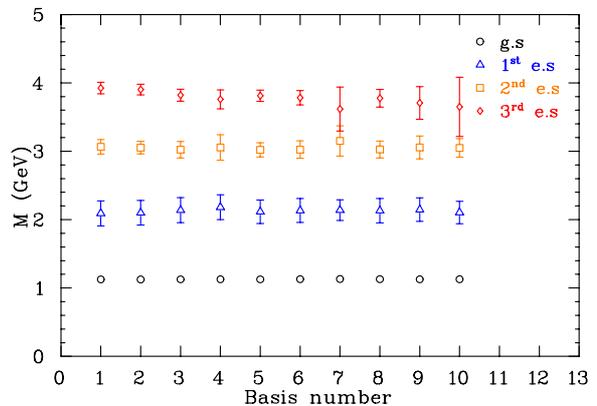} 
  \end{center}
\vspace{-0.6cm}
    \caption{(Color online). Masses of the $N{\frac{1}{2}}^{+}$ energy
      states for various $4\times 4$ correlation matrix bases as given in
      Table~\ref{table:varieties_of_4x4_matrices}, for
      $\kappa_{ud}=0.13770$, $m_{\pi}=293\,
        \text{MeV}$, over 50 configurations.}
 \label{fig:4x4_x1x1_Kud01377000Ks01364000}
\end{figure}

The consistency of the extracted masses from all the $4\times 4$
matrices considered in Table~\ref{table:varieties_of_4x4_matrices} is
illustrated in Fig.~\ref{fig:4x4_x1x1_Kud01377000Ks01364000}. In
particular, the ground and Roper states are robust.
Both lower and higher smearing radii are beneficial for spanning the
space of states at all quark mass. However, we avoid bases which
include extreme smearing counts (400 and 800) as these often provide
ill-defined correlation matrices. Hence, we select basis number 3 as the
focus of subsequent analysis.

\begin{figure*}[t]
  \begin{center}
 \includegraphics[height=0.8\textwidth,angle=90]{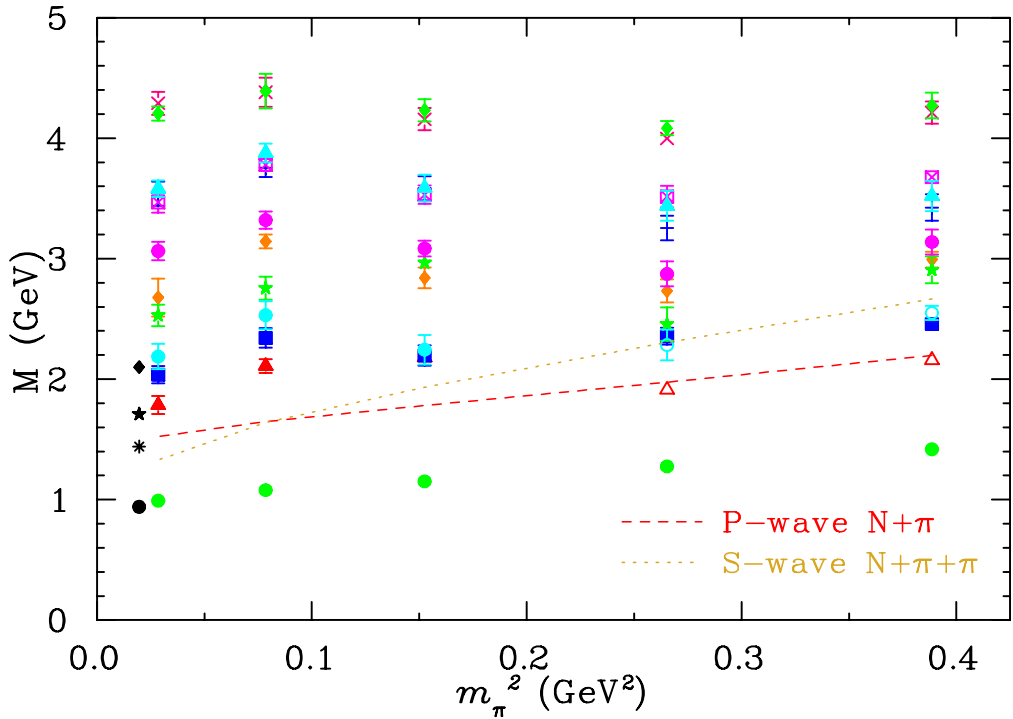}
  \end{center} 
\vspace{-0.5cm}
    \caption{(Color online) Masses of the positive-parity
       states of the nucleon. Physical values are obtained
       from Ref.~\cite{Amsler:2008zzb} and plotted at the far left.
       We note the presence of three approximately degenerate states
       for the Roper at the third quark mass which split at the
       lighter quark masses.
       Lattice results for the Roper (red triangles) reveal
       significant chiral curvature towards the physical
       mass. Non-interacting $P$-wave 
       and $S$-wave threshold states are also presented.}
\label{fig:m.12x12}
\vspace{-0.2cm}
\end{figure*}

In Fig.~\ref{fig:m.12x12}, results for the first 12 eigenstate
energies are reported for the five quark masses available.  The scale
is set via the Sommer parameter \cite{Sommer:1993ce}.  The 12 states
are drawn from three $8\times 8$ correlation-matrix analyses for pairs
of $\chi_1$, $\chi_2$ and $\chi_4$.  The matrices are formed with each
interpolator having four levels of smearing.  Whereas the $\chi_1$,
$\chi_2$ and $\chi_4$, $\chi_2$ analyses reveal the same spectrum,
four new states are revealed in the $\chi_1$, $\chi_4$ analysis in
place of the four states dominated by $\chi_2$.  

The colour coding and symbols for the states of Fig.~\ref{fig:m.12x12}
illustrate the flow of the states from one quark mass to the next as
identified by the eigenvector annihilating the state.  The structure
of the eigenvectors isolating the states varies only slowly from one
quark mass to the next making it easy to trace the propagation of the
states from the heavy to the light quark-mass region.  We note that
this analysis reveals the same number of energy states as in the
physical resonance spectrum below 2.2 GeV~\cite{Amsler:2008zzb}.

Among the most significant results of this investigation is the manner
in which the extracted Roper state (filled triangles) approaches the
physical value.  The significant curvature in the chiral regime
indicates the important role played by mesonic dressings of the
Roper. Non-interacting $P$-wave $N\pi$ ($E_{N}+E_{\pi}$) and $S$-wave
$N\pi\pi$ ($M_{N}+M_{\pi}+M_{\pi}$) threshold scattering states are
presented by the dashed and dotted lines, respectively.  For the two
large quark masses, the results sit close to the $P$-wave $N \pi$
scattering threshold whereas the masses for the lighter three quark
masses sit much higher.  While there is some evidence for the $P$- and
$S$-wave scattering states for the two heavier quarks, there is no
evidence of these states at light quark masses.

A possible explanation for this feature is that the attractive
mass-dependent and spin-dependent forces which are necessary for the
formation of a strong resonance only have sufficient strength at light
quark masses.  For example, it is typical to encounter spin-dependent
forces which are inversely proportional to the product of the quark
masses undergoing gluon exchange.  
As there is no evidence for the threshold scattering states with
back-to-back momenta of one lattice unit, $\vec p = (2 \pi/ L_x, 0,
0)$, at the lightest three quark masses, we find it unlikely that
scattering states with the next back-to-back momenta, $\vec p = (2
\pi/ L_x, 2 \pi/ L_y, 0)$ would suddenly appear in our spectrum.
At light quark masses, resonant eigenstates dominated by
single-particle states dominate the spectral function whereas at heavy
quark masses, only the multiparticle states have spectral strength
sufficient to be seen in the spectrum.

In addition to this quark-mass effect, there is a well-known volume
effect.  The couplings to the multi-particle meson-baryon states are
suppressed by $1/\sqrt{V}$ relative to states dominated by a
single-particle state.  On our relatively large volume, it is likely
that multi-particle states will be suppressed and missed in our
spectrum, particularly at lighter quark masses where the quark-mass
effect also acts to suppress the spectral strength.  Further analysis
of finite volume effects~\cite{Young:2002cj} on the spectrum is
desirable.

Future calculations should also investigate the use of five-quark
operators to ensure better overlap with the multi-particle states.
This type of novel work using the stochastic LapH method is in
progress~\cite{Morningstar:2011ka}.  Indeed, complete knowledge of the
spectrum is required for a definitive determination of the properties
of the Roper resonance.

\begin{figure}
  \begin{center}
 \includegraphics[height=0.46\textwidth,angle=90]{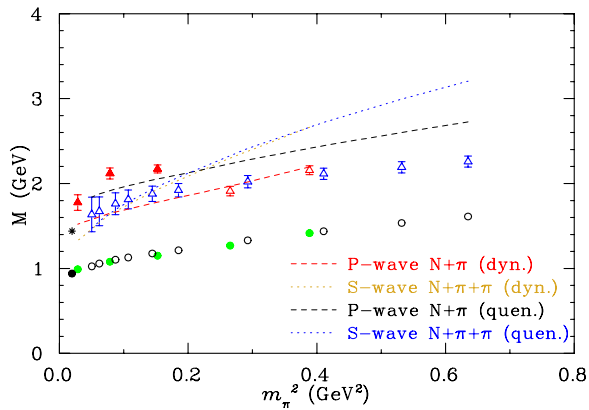}
  \end{center} 
\vspace{-0.5cm}
    \caption{(Color online).  A comparison of the low-lying
      positive-parity spectrum of dynamical QCD (full symbols) and
      quenched QCD results (open symbols) from
      Ref.~\cite{Mahbub:2009aa}. In the quenched case, the lattice volume
      was ${(2\, \rm{fm})^{3}}$.}
\label{fig:m.4x4_dyn_quen_pswave}
\vspace{-0.2cm}
\end{figure}

Fig.~\ref{fig:m.4x4_dyn_quen_pswave} provides a comparison
of our results in full QCD with earlier results in quenched QCD
\cite{Mahbub:2009aa},
 where the effects of dynamical quark loops are not considered.  
While the ground state of the nucleon in quenched (open
symbols) and full QCD (full symbols) are in reasonable agreement,
significant differences are observed for the Roper in the light quark
mass regime. This is an important discovery emphasizing
the  role of dynamical fermion loops in the structure of the
Roper. Note, the difference between the $P$-wave $\pi N$ threshold scattering
state energies in full and quenched QCD is due to the difference in the
lattice volume. 

We have also compared~\cite{Mahbub:2011rm} our extracted nucleon
spectrum with the HSC collaboration's~\cite{Edwards:2011jj} results at
${m_{\pi}}^{2}\approx 0.27$ ${\text{GeV}}^{2}$ where multi-particle
states are seen in our analysis. Taking into account their
small-volume of $(1.97\, \text{fm})^{3}$, which shifts the energy of
the $P$-wave scattering states for example, both spectra are in good
qualitative agreement.  Of particular note is the identification of
the same number of energy states below 3 GeV.

This investigation is the first to illustrate the manner in which the
Roper resonance of Nature manifests itself in today's best numerical
simulations of QCD.  The quark mass dependence of the state revealed
herein substantiates the essential role of dynamical fermions and
their associated non-trivial light-mesonic dressings of baryons, which
give rise to significant chiral non-analytic curvature in the Roper
mass in the chiral regime.

This research was undertaken on the NCI National Facility in Canberra,
Australia, which is supported by the Australian Commonwealth
Government. We also acknowledge eResearch SA for generous grants of
supercomputing time.  This research is supported by the Australian
Research Council.

\end{document}